\documentclass[conference]{IEEEtran}
\IEEEoverridecommandlockouts

\usepackage[noadjust]{cite}
\usepackage{hyperref}
\usepackage{amsmath,amssymb,amsfonts}
\usepackage[ruled,vlined]{algorithm2e}
\usepackage{graphicx}
\usepackage{textcomp}
\usepackage{siunitx}
\usepackage{lscape}
\usepackage[table]{xcolor}
\usepackage{tabularx}
\usepackage{rotating}
\usepackage{multicol}
\usepackage{subcaption}
\usepackage{bm}
\usepackage{float}
\usepackage{multirow}

\makeatletter
\def\endthebibliography{%
	\def\@noitemerr{\@latex@warning{Empty `thebibliography' environment}}%
	\endlist
}
\makeatother

\DeclareSIUnit\bcn{beacon}

\begin{document}

\title{Authenticated time for detecting GNSS attacks}

\author{
	\IEEEauthorblockN{Marco Spanghero}
	\IEEEauthorblockA{\textit{Networked Systems Security Group} \\
		\textit{KTH Royal Institute of Technology}\\
		Stockholm, Sweden \\
		marcosp@kth.se}
	\and
	\IEEEauthorblockN{Kewei Zhang}
		\IEEEauthorblockA{\textit{Networked Systems Security Group} \\
		\textit{KTH Royal Institute of Technology}\\
		Stockholm, Sweden \\
		kewei@kth.se}
	\and
	\IEEEauthorblockN{Panagiotis Papadimitratos}
	\IEEEauthorblockA{\textit{Networked Systems Security Group} \\
		\textit{KTH Royal Institute of Technology}\\
		Stockholm, Sweden \\
		papadim@kth.se}
	
}

\maketitle

\begin{abstract}
Information cross-validation can be a powerful tool to detect manipulated, dubious GNSS data. A promising approach is to leverage time obtained over networks a mobile device can connect to, and detect discrepancies between the GNSS-provided time and the network time. The challenge lies in having reliably both accurate and trustworthy network time as the basis for the GNSS attack detection. Here, we provide a concrete proposal that leverages, together with the network time servers, the nearly ubiquitous IEEE 802.11 (Wi-Fi) infrastructure. Our framework supports application-layer, secure and robust real time broadcasting by Wi-Fi Access Points (APs), based on hash chains and infrequent digital signatures verification to minimize computational and communication overhead, allowing mobile nodes to efficiently obtain authenticated and rich time information as they roam. We pair this method with Network Time Security (NTS), for enhanced resilience through multiple sources, available, ideally, simultaneously.
We analyze the performance of our scheme in a dedicated setup, gauging the overhead for authenticated time data (Wi-Fi timestamped beacons and NTS). The results show that it is possible to provide security for the external to GNSS time sources, with minimal overhead for authentication and integrity, even when the GNSS-equipped nodes are mobile, and thus have short interactions with the Wi-Fi infrastructure and possibly intermittent Internet connectivity, as well as limited resources. 
\end{abstract}

\section{Introduction}
\label{section:intro}
Cost effective, compact, high precision \textit{Global Navigation Satellite Systems} (GNSS) receivers enable the integration of precise location and timing services in diverse, mobile platforms, ranging from cars to smart-phones to Internet of Things devices. While bringing extensive benefits to users, the intrinsic lack of security in GNSS \cite{volpe2001vulnerability} exposes new attack surfaces. Civilian-grade GPS and GLONASS do not guarantee the authenticity of the received signals. Even though this problem is being addressed by GALILEO's Navigation Message Authentication (NMA) (\cite{european2015,fernandez2016navigation,walker2015galileo}), it will not preclude all the attacks (e.g., Distance Decreasing \cite{zhang2015gnss,zhang2019on}). At the same time, easy accessibility to software-defined radios and open source software simulators, reduce the cost of platforms capable of mounting advanced attacks (\cite{zeng2018all, Schmidt2016, Zeng2017, Huang2015}). This is a significant threat, successfully demonstrated in simulated attack scenarios and it can disrupt the availability, accuracy or correctness of the GNSS information. 

This is why defensive methods have been investigated. Beyond modifying GNSS, e.g. with NMA, various defense mechanisms can be implemented on the receiver side (\cite{papadimitratos2012method, psiaki2016gnss, PapadimitratosJa:C:2008, ZhangP:C:2019b, ZhangP:C:2019c}). Countermeasures based on Inertial Measurement Units (IMUs) (\cite{curran2017use, khanafseh2014gps}), signal processing such as Automatic Gain Control (AGC) monitoring and phase measurements across multiple antennas, are few examples of defenses based on integrating external sensors or exploiting physical characteristics of the signal (\cite{montgomery2009receiver, psiaki2013gnss, bastide2003automatic, borowski2012detecting, psiaki2014gnss}).

Most of the developed defense approaches do not leverage the likely network connectivity of the GNSS equipped platforms. In fact, modern devices can have access to several sources of information for validating GNSS obtained location and timing. Such a capability can be valuable as it could be readily deployed to safeguard GNSS-based position (and timing). The recent results in \cite{kzmsppPLANS2020} exploit the co-existence of multiple time sources to cross-validate time information obtained from the GNSS receiver, demonstrating how a combination of different time sources can be used to detect deviations in GNSS time and detect GNSS attacks that caused such time deviations. In the context of protecting mobile phones from spoofing, side-information based approaches, such as time sources, inertial sensors and A-GPS (Assisted GPS) information \cite{8373546} are also investigated. 

Even though existing approaches, such as \cite{kzmsppPLANS2020}, raise the bar, it is paramount to have trustworthy information from the leveraged alternate sources. If not, in the presence of an adversary that can attack both the GNSS side (e.g., spoofing) and the Internet side (e.g., impersonating or manipulating the alternate source of time information), the detection would fail. Time sources such as the Network Time Protocol (NTP), 3G/4G Radio Base Stations (RBSs) and WiFi Access Points (APs) can provide accurate timestamps. But they do not immediately offer protection of the time information, allowing powerful attackers to feed the victim receiver with tampered time information. The lack of authentication in NTP frames and WiFI beacons can be exploited by a malicious actor that forges messages to arbitrarily skew the time of requesting clients; it can block legitimate replies and substitute them with bogus ones \cite{Malhotra2017}, aiming to overtake the attack detection system.

Beyond the aforementioned vulnerability, using NTP and Internet-based time services assumes the GNSS-equipped device is able to frequently connect to the remote infrastructure. In high mobility scenarios, the interaction between different entities is limited in time by their relative speed; establishing a link with an AP, ideally a trusted one, might be hard for highly mobile nodes. Cellular internet access can be more effective, yet it can be challenging in dense network areas or in rural environments. Cellular RBS physical layer can provide accurate time information (Primary Synchronization Signals (PSS) \cite{3gpp200305,3gpp201605}), but exploiting this information comes at a significantly higher implementation complexity for the receiver. Furthermore, as the cellular infrastructure relies on GNSS to maintain synchronization between RBSs, an RBS may be under the influence of an attacker we want to defend the mobile nodes from. Hence, a connection-less approach, based on opportunistic information could be less complex to implement and applicable, particularly in an urban environment, with dense APS deployment. 

It is, of course, critical to authenticate the timing information from the time servers and the APs. To address this, we investigate the use of Network Time Security (RFC 7380 \cite{ietf-ntp-using-nts-for-ntp-28}) and we propose an authentication scheme for WiFi beacons, studying different configurations for authenticated time broadcast. 
The result is a framework that supports authenticated time provision to cross-validate GNSS-obtained time, leveraging mainstream mobile systems.
Building on the approach of  \cite{kzmsppPLANS2020} and \cite{8373546}, our scheme supports different options to safeguard time information. In particular, we investigate how to achieve this with low computation and communication cost, and reasonable credential management complexity. We are interested in a solution that, while being easy to deploy, remains effective even with intermittent Internet connectivity. Existing numerous timing servers, ubiquitous, redundant WiFi APs for resilience to misbehavior, and low cost cryptographic protection are the ingredients of our proposal. 

The paper is organized as follows: Section \ref{section:system-model} presents the system and attacker model, Section \ref{section:scenarios} discusses how to safeguard external time information towards detecting GNSS attacks. Section \ref{section:experimental} presents implementation details and the computational overhead of our proposal. Section \ref{section:conclusion} discusses future developments and concludes.
\section{System and Attack Model}
\label{section:system-model}
We consider a mobile client that interacts with a fixed terrestrial network and obtains GNSS information through a GNSS receiver. Clients can access Internet over WiFi APs, possibly authenticated. Clients receive broadcast messages from APs within range, without necessarily being associated with those APs. The APs advertise WiFi networks with 802.11 protocol compliant beacons and in a single hop network, where all the receivers obtain a wireless broadcast message (beacon) practically simultaneously.

Typically, GNSS attackers in the literature influence victim GNSS receivers. We extend this model and consider an adversary that can also attack the terrestrial network(s) the GNSS-enabled devices connect to. The attacker can attempt to manipulate the alternative sources of information (WiFi beacons and NTP frames) in a coordinated effort towards deceiving the GNSS receiver, notably towards defeating the GNSS cross-validation approach. 

We do not limit the capacity and objectives of the attacker in terms the deviations it can cause to the GNSS Position Velocity and Timing (PVT) solution. In addition, we do not presume which specific methods can be used by the adversary.

Let \bm{$\rho$} be a \textit{n-vector} of pseudo-ranges, \bm{$H$} be a \textit{n x 4 matrix} of receiver state observations and \bm{$x$} be the PVT solution for the corresponding observation, nominally $\bm{x} = [x,y,z,T_{GNSS}]$, $ \bm{v}$, \textit{n x 1 vector}, be a measurement noise. Let $\bm{f}$ be an \textit{n x 1 vector} that corresponds to the adversarial action; then the following models a generic GNSS receiver under adversarial influence:
 $$\bm{\rho} = \bm{H} \cdot \bm{x} + \bm{f} + \bm{v}$$
Adversaries able to craft $\bm{f}$ to selectively tamper only with the location, or only with the velocity, part of the PVT solution, while intentionally not modifying the time information, cannot be thwarted by the method considered here or by any other time-based spoofing detection method \cite{Humphreys}.

The attacker, provided with WiFi and, possibly, other radios, and Internet access can attempt forging 802.11 beacon frames. The content of the beacon frames can be either malicious (containing misleading timing information) or meaningless, to consume resources of the receiver. The attacker can spoof or meacon single beacons or sequences of beacons and NTP frames to mask the misbehavior in the GNSS information. 

The attacker and the victim can be subject to the same propagation delay when receiving messages from an AP. Furthermore, the adversary may have higher computational power than the clients and the APs, but she cannot obtain system credentials and cryptographic keys.
\section{Safeguarding Time-based GNSS verification}
\label{section:scenarios}

\begin{figure}[h]
	\centering
	\includegraphics[width=\linewidth]{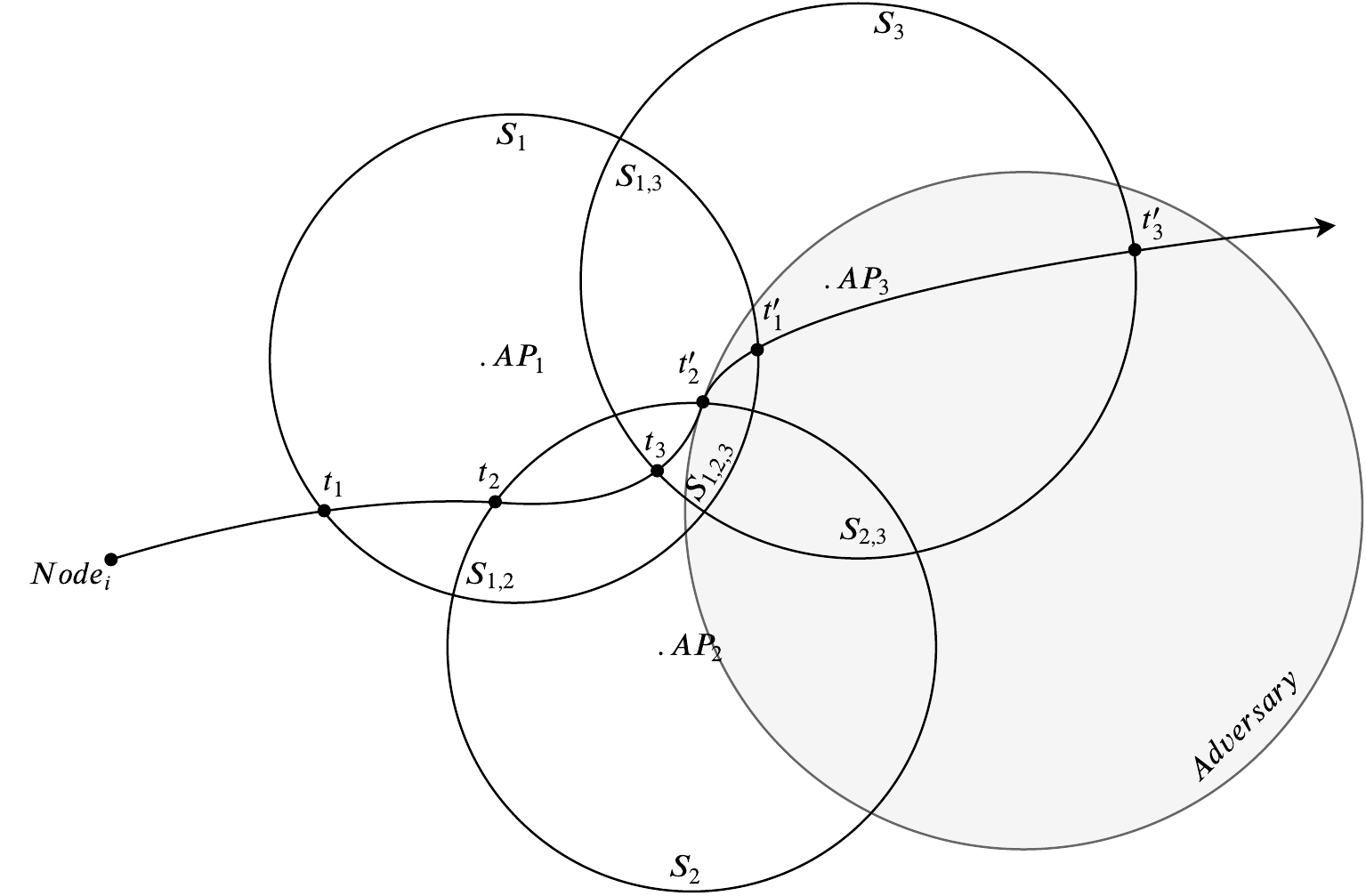}
	\caption{Mobile node transversing several AP coverage zones. One node can be in range of multiple APs at the same time. The adversarial zone is influenced by the combined effort of an attacker spoofing GNSS receiver while disseminating bogus beacon information, or even introducing rogue APs.}
	\label{fig:transverse-adversary}
\end{figure}

Augmentation information from externally provided time references can be useful only under the assumption that the terrestrial networking infrastructure is secure. Consider Figure~\ref{fig:transverse-adversary}, where an adversary controls part of the infrastructure and can emulate/forge bogus timing information, in a coordinated effort in parallel to the GNSS attack. Within the attacker's influence area, mobile clients are subject to the GNSS spoofer actions and they receive a combination of legitimate and malicious time information, without being able to distinguish the former from the latter. 

Wi-Fi APs can offer secure links to mobile clients in range and to the Internet, provided that the client has enough time to establish a connection and that it has the necessary credentials (i.e., Wi-Fi password, or certificate). Such a link can be leveraged to access Internet time services, to cross-check the GNSS time with that provided by the Network Time Protocol (NTP) infrastructure. If the attacker can influence only the GNSS receiver of the receiver, this approach is valid and can detect adversaries that tamper with the GNSS, if the modification of the GNSS time is larger than the accuracy provided by NTP. An advanced attacker that hijacks the NTP server connection could mask the adversarial action on the GNSS receiver by forging bogus NTP replies. This is possible even if the connection between the AP and the mobile client is secured, assuming the attacker knows which NTP servers are used by the client and the interaction between the client and the NTP server is unauthenticated. In case an attacker is able to tamper with the NTP information, Network Time Security protocol (NTS) can be used to provide authenticated NTP frames.

Although this raises the bar for a potential attacker, there are limitations due to the structure of the Wi-Fi protocol. As an example, mobile clients are able to connect to only one AP at the time. There are instances when the mobile device is not able to connect to any remote time provider. Clearly, the mobile node can connect to the APs it has valid credentials for. Furthermore, Wi-Fi associations are slow and the authentication process introduces latency. Depending on the mobility of the client, it might be infeasible for it to establish a secure connection. 

In such situations, with limited (secure) access to the Internet, the Wi-Fi infrastructure can still be useful, in an opportunistic manner. The client does not connect to the APs but it exploits the management and advertisement information distributed by the APs, more specifically Wi-Fi beacons, which provide a rich source of timing information. Defined in Section 9.3.3 of the IEEE Wireless LAN MAC Specification \cite{6178212} as management frames, they are used to advertise networks and AP capabilities. Additionally, they contain a 64-bit, microsecond resolution timestamp that can be used to synchronize devices within the range of the AP; providing a precise source of relative time information. 

On the other hand, as beacons are not authenticated, based on the current standard and deployment status they provide a source of time of low trustworthiness. It is relatively simple for an attacker to spoof/meacon existing AP beacons or emulate new APs with timestamps that can support the adversarial action on the GNSS side. This would still require a sophisticated adversary to orchestrate a coordinated attack on the GNSS receiver and the terrestrial infrastructure. Interestingly, this might be comparably easier than hijacking the connection to an NTP server. To leverage unauthenticated Wi-Fi beacons in an attempt to validate the GNSS-provided time, the mobile client might collect information from a diverse set of APs. This comes at the cost of elevated overhead and would require assumptions on the maximum number of APs the adversary is capable of controlling/emulating. Overall, a scheme difficult to be effective. 

Authenticated timestamps would provide an additional layer of security for the GNSS attack detection. Several options can be considered. If the mobile node can associate itself with the AP, it could use a shared symmetric key to authenticate beacons. Establishing such a symmetric key can be done in several standard ways, but it is subject to the latency of establishing a secure connection with the AP. Moreover, even so, it would allow any of the mobile nodes to transmit such beacons.

The APs can authenticate broadcast beacons. A naive solution would be to use asymmetric/public key cryptography to digitally sign every message/beacon. The creation and verification of a digital signature for each beacon would introduce significant overhead due to the relatively high beacon rate ($\approx$ \SI{10}{\bcn \per \second}). This would be costly for the APs and not sustainable for constrained mobile devices, enabling clogging or resource depletion Denial of Service (DoS) attacks~\cite{JIN2019101775}.

\subsection{Efficient beacon authentication}
\label{section:approach}
The approach used here is to predominantly apply symmetric key cryptography for broadcast authentication with the necessary and infrequent use of asymmetric key digital signatures. An approach such as \cite{DBLP:conf/ndss/PerrigCST01} provides efficient broadcast authentication. To achieve lightweight authentication of beacons (and timestamps) here, the hash chain elements are used as symmetric keys. Each beacon is protected by an authentication code ($MAC^{n}$) calculated with the hash chain element that will be disclosed at the following transmission cycle, the subsequent beacon. Upon receipt of that latter beacon, the mobile nodes in range can authenticate the previous, sufficiently early received, beacon. Beacons very close to (and trivially after) the key disclosure are rejected, given the single hop broadcast. 

\begin{figure}[ht]
 	\centering
	\includegraphics[width=\linewidth]{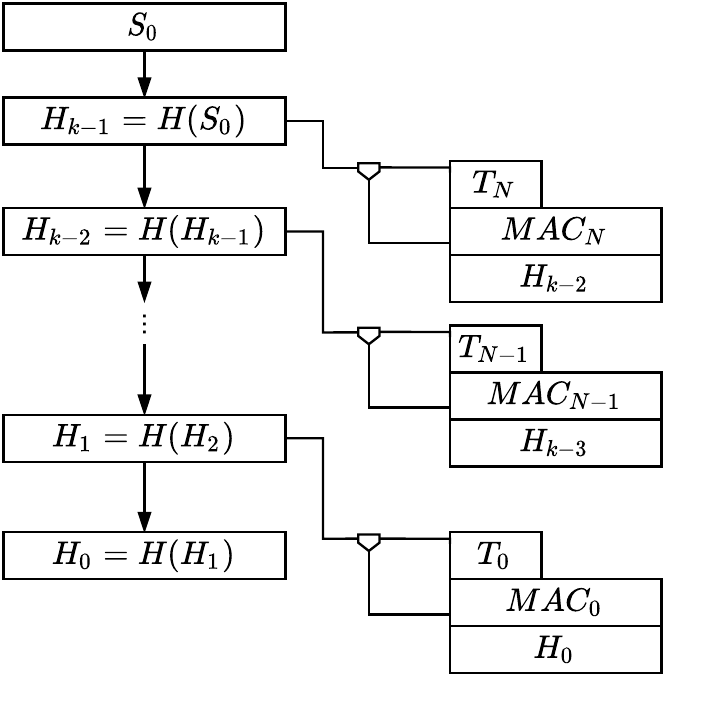}
	\caption{Chain structure and generated payloads. Infrequent digital signatures are applied to the hash chain elements}
	\label{figure:modified-chains}
 \end{figure}

Fig. \ref{figure:modified-chains} shows the generation of the hash chain and the order of the use of its elements. The last element, the anchor, is digitally signed by the AP, to bind a hash chain to the specific AP. This essentially allows any client that moves within the AP range to obtain a digitally signed anchor for the hash chain and validate any subsequent beacon authentication code. The public key that corresponds to the private key, used for digital signing, is certified by a trusted third party, a Certification Authority or Public Key Infrastructure (PKI). For simplicity, here, each AP identity is bound to its public key. The resultant AP certificate can either be made available over the Internet or it can be periodically broadcast by the AP, attached to a subset of the transmitted beacons (leveraging an idea from ~\cite{CalandrielloPHL:J:2011, JinP:C:2016}).

The hash chain is pre-computed by the AP and it can cover long periods of time. Due to the high frequency of the beacons and given that each hash chain element is used as a one-time key material (thus requiring one element per each beacon timestamp), the length of the hash chain can be significant. As mobile clients can join the coverage area of an AP at any point in time, if the client is provided only with the initial anchor to the hash chain, it might be forced to perform a large number of hash calculations before confirming the validity of any hash chain element and message. The AP can digitally sign intermediate anchors, to be disclosed by the AP as the usage of the hash chain progresses. By verifying the digital signature of the intermediate anchor, the mobile client calculates a much lower number of hashes initially. Different strategies are possible to distribute anchors, similarly to the distribution of the AP certificates.

We can exploit the regularity of the beacon emission. The hash chain is generated in advance, alongside with the signed anchors. MACs are computed and packets are assembled and scheduled for transmission at precise times, allowing the AP to execute all the cryptographic operations outside the transmission critical path. 

Finally, a reminder that the scheme does not operate in isolation but together with possibly less frequent access to internet time servers. This provides a time reference to mobile nodes, useful for this broadcast authentication as well as the ability to detect replays of sequences of authenticated beacons. To be more effective against such attacks, additional information on the AP and the network context can be used.

\section{Test environment Setup and results}
\label{section:experimental}

\begin{figure}[ht]
	\centering
	\includegraphics[width=\linewidth]{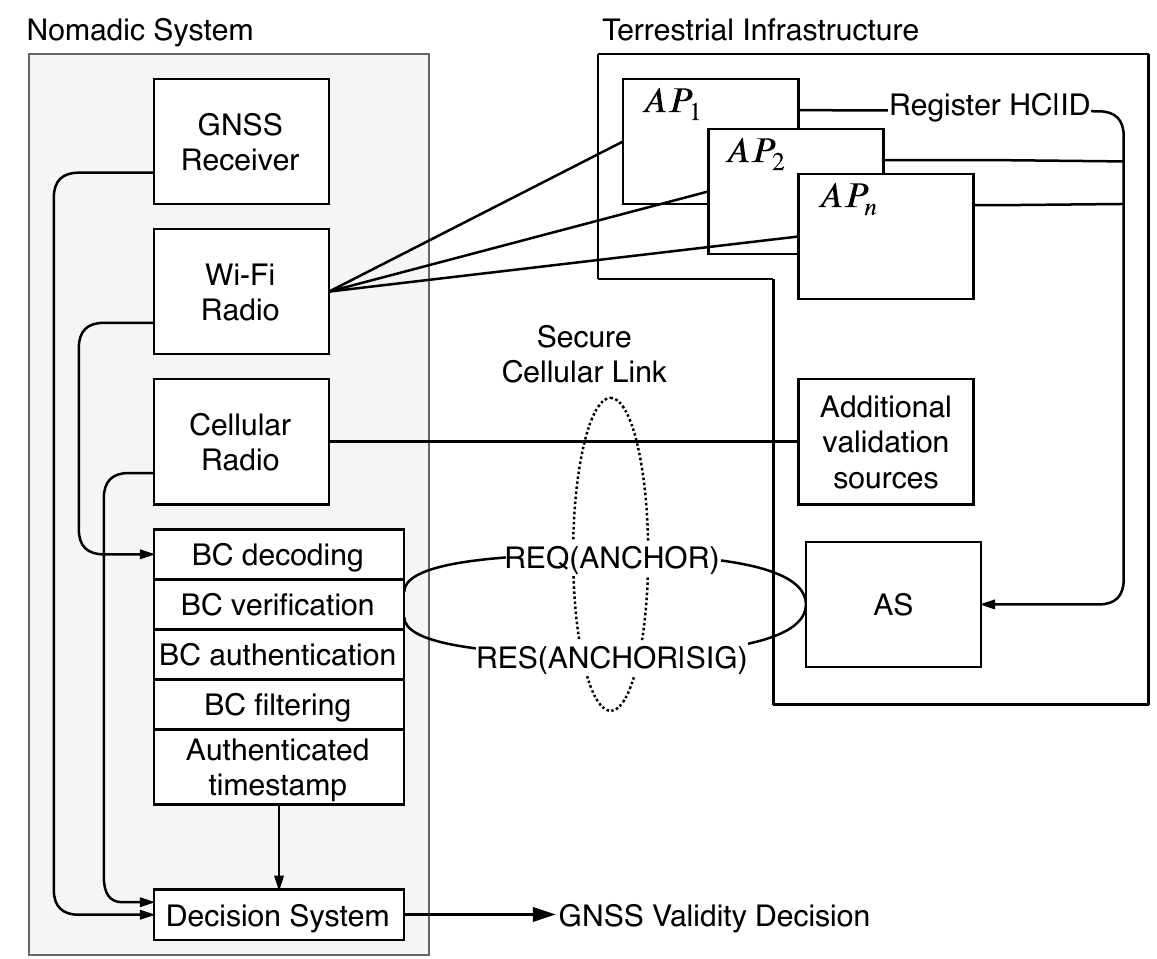}
	\caption{Framework for secure time broadcasting and Wi-Fi beacons authentication. }
	\label{fig:system-arch}
\end{figure}

The method described in Section \ref{section:approach} is implemented in a laboratory test setup. The authentication scheme for the Wi-Fi beacons allows mobile clients to detect APs, obtain digitally signed anchors and verify the beacons payload. Figure~\ref{fig:system-arch} presents the implemented system architecture and the interaction of its components. The authentication scheme is implemented at the application layer and we evaluate a deployment of the system with one client, one AP (supporting our authentication scheme) and several APs without authentication.  We pre-install the certificate of the AP in the client. This does not affect the generality of the setup, as certificates and anchors can be broadcasted.

The test system is composed of the following components:
\begin{itemize}
	\item Host machine: Intel I7 CPU running a Linux system whose kernel supports high precision timing
	\item Host Wi-Fi card: Intel Corporation Wireless-AC 9260
	\item Host virtual Wi-Fi card: mac80211\_hwsim kernel module
	\item Wi-Fi beacons: from surrounding APs in the test area
	\item Secure Wi-Fi beacons: dedicated beacon emulator in the test area
	\item NTP servers: three servers in Sweden (provided by Netnod \cite{netnod-ntp})
	\item NTS servers: three servers in Sweden (provided by Netnod \cite{netnod-ntp})
	\item GPS PVT rate: \SI{1}{Hz}
	\item Ublox EVK-6T evaluation kit \cite{ublox2012}
\end{itemize}

To emulate a mobility scenario, the client interacts with the AP by receiving the beacon broadcast information over Wi-Fi, in a non connected state. Additionally, it communicates with the NTS/NTP infrastructure, over a wireless link. 

We developed a setup that allows to emulate a beacon transmitter. We rely on high accuracy software timers to emit beacons at a constant rate of 100TU. To transmit custom Wi-Fi beacon frames, the wireless cards are configured in "monitor" mode. This allows leveraging the packet injection capabilities of the wireless interface to send raw frames without directly interacting with the driver. To achieve the highest accuracy, it would be necessary to modify the hardware driver for the Wireless interface. To the best of our knowledge, in the current 802.11 implementation in Linux, beacons are configured in user space as static templates. The information is received in the kernel module where channel state information is added and ultimately it is scheduled for transmission by the driver, where the final timestamp is added. This static structure does not allow directly supporting our scheme, as information contained in the Wi-Fi beacons needs to be updated at each transmission cycle. Nevertheless, there is no limitation with the current standard to have dynamic beacon payloads, provided that the compulsory information is included. 

To test the concept, the \textit{mac80211\_hwsim} Wi-Fi simulation driver is used. This kernel module allows setting a virtual Wi-Fi devices that can either be configured as a Station (STA) or Access Point (AP). The virtual Wi-Fi device supports packet injection and behave like ideal Wi-Fi devices. The major limitation is that it does not take into account real propagation medium or coexistence with other wireless networks. The generation, transmission and detection of the beacons uses Scapy framework \cite{scapy} and is Python3.7 based.

The specialized authentication payload is included in the Wi-Fi beacon structure, at the Vendor Specific tag fields. This is allowed by the protocol, as long as the assembled packet is smaller than the maximum size (2320 octets \cite{6178212}).
Timestamps are placed in the dedicated mandatory field of the beacon consisting of 8 octets, corresponding to a 64 bit microsecond timestamp.
All clients that do not support the Vendor Specific fields we define, simply ignore them; our implementation can coexist with existing wireless networks without disrupting their behavior. The implementation does not depend on a specific signature method and it could be extended to support different signatures. 

The integration of Wi-Fi APs with authenticated NTS servers provides a solid basis to detect time-shifting GNSS spoofing attacks and more generally assesses the health of the GNSS information. We evaluate the potential overhead introduced by the authentication of the NTP information in NTS and by our time authentication scheme, to understand whether the same accuracy as the non authenticated options is achievable. We expect that authentication causes minimal, if any, degradation.

\subsection{Evaluation results}
No substantial performance degradation was observed while using NTS for retrieving absolute timestamps from a remote server over 4G networks when compared to NTP in the same conditions. Table \ref{table:ntsperf} shows the results of a 6-hour test against three NTS servers provided by NetNod in Sweden. Although there is a small degradation in the offset of the synchronizing client when using NTS, the standard deviation of RTD and offset of the two methods are compatible, and so is their accuracy.

\begin{table}[h!]
 \begin{tabularx}{\columnwidth}{||X | X | X||} 
 \hline
 Source & RTD & Offset \\
 \hline
 \hline
 \multirow{2}{4em}{nts.sth2.ntp.se \\ 2a01:3f7:2:202::201} &$\mu: 2.825\mathrm{e}{-2}$ & $\mu: 1.228\mathrm{e}{-3}$\\ & $\sigma: 7.200\mathrm{e}{-3}$ & $\sigma: 4.331\mathrm{e}{-3}$\\
 \hline
 \multirow{2}{4em}{nts.sth1.ntp.se \\ 2a01:3f7:2:52::10} &$\mu: 2.765\mathrm{e}{-2}$ & $\mu: 9.380\mathrm{e}{-4}$\\ & $\sigma: 7.151\mathrm{e}{-3}$ & $\sigma: 3.309\mathrm{e}{-3}$\\
 \hline
 \multirow{2}{4em}{nts.sth.ntp.se \\ 2a01:3f7:2:62::10} &$\mu: 2.774\mathrm{e}{e-2}$ & $\mu: 1.042\mathrm{e}{-3}$\\ & $\sigma: 6.970\mathrm{e}{-3}$ & $\sigma: 3.131\mathrm{e}{-3}$\\
 \hline 
 \end{tabularx}
\caption{Performance of NTS over cellular 4G connection (measurements in seconds (\si{\second}))}
\label{table:ntsperf}
\end{table}

\begin{table}[h!]
 \begin{tabularx}{\columnwidth}{||X | X | X||} 
 \hline
 Source & RTD & Offset \\
 \hline
 \hline
 \multirow{2}{4em}{sth1.ntp.se \\ 2a01:3f7:2:1::1} &$\mu: 2.673\mathrm{e}{-2}$ & $\mu: 3.177\mathrm{e}{-4}$\\ & $\sigma: 6.512\mathrm{e}{-3}$ & $\sigma: 3.409\mathrm{e}{-3}$\\
 \hline
 \multirow{2}{4em}{sth2.ntp.se \\ 2a01:3f7:2:2::1} &$\mu: 2.631\mathrm{e}{-2}$ & $\mu: 4.223\mathrm{e}{-4}$\\ & $\sigma: 5.526\mathrm{e}{-3}$ & $\sigma: 2.975\mathrm{e}{-3}$\\
 \hline
 \multirow{2}{4em}{npt3.sptime.se \\ 2001:6b0:42:3::123} &$\mu: 3.931\mathrm{e}{-2}$ & $\mu: -5.802\mathrm{e}{-3}$\\ & $\sigma: 1.375\mathrm{e}{-3}$ & $\sigma: 6.985\mathrm{e}{-3}$\\
 \hline 
 \end{tabularx}
\caption{Performance of NTP over cellular 4G connection (measurements in (\si{\second}))}
\label{table:ntpperf}
\end{table}

\begin{figure*}[!t]
	\centering
	\begin{subfigure}{0.49\textwidth}
		\centering
		\includegraphics[height=2.5in]{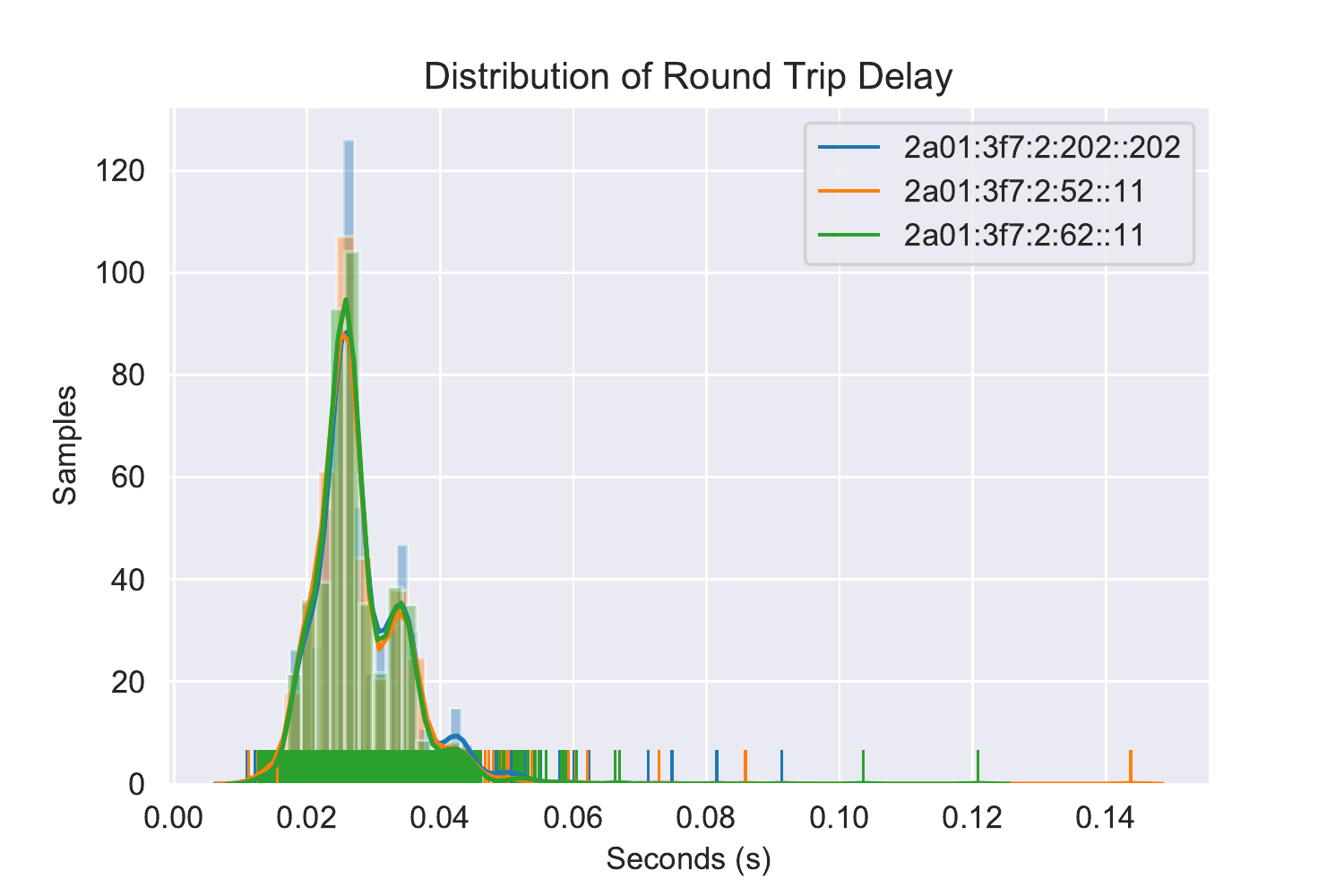}
		\caption{Histogram of NTS Round Trip Delay distribution for three NTS servers in Sweden. Measurement details are provided in Table \ref{table:ntsperf}}
	\end{subfigure}
	~
	\begin{subfigure}{0.49\textwidth}
		\centering
		\includegraphics[height=2.5in]{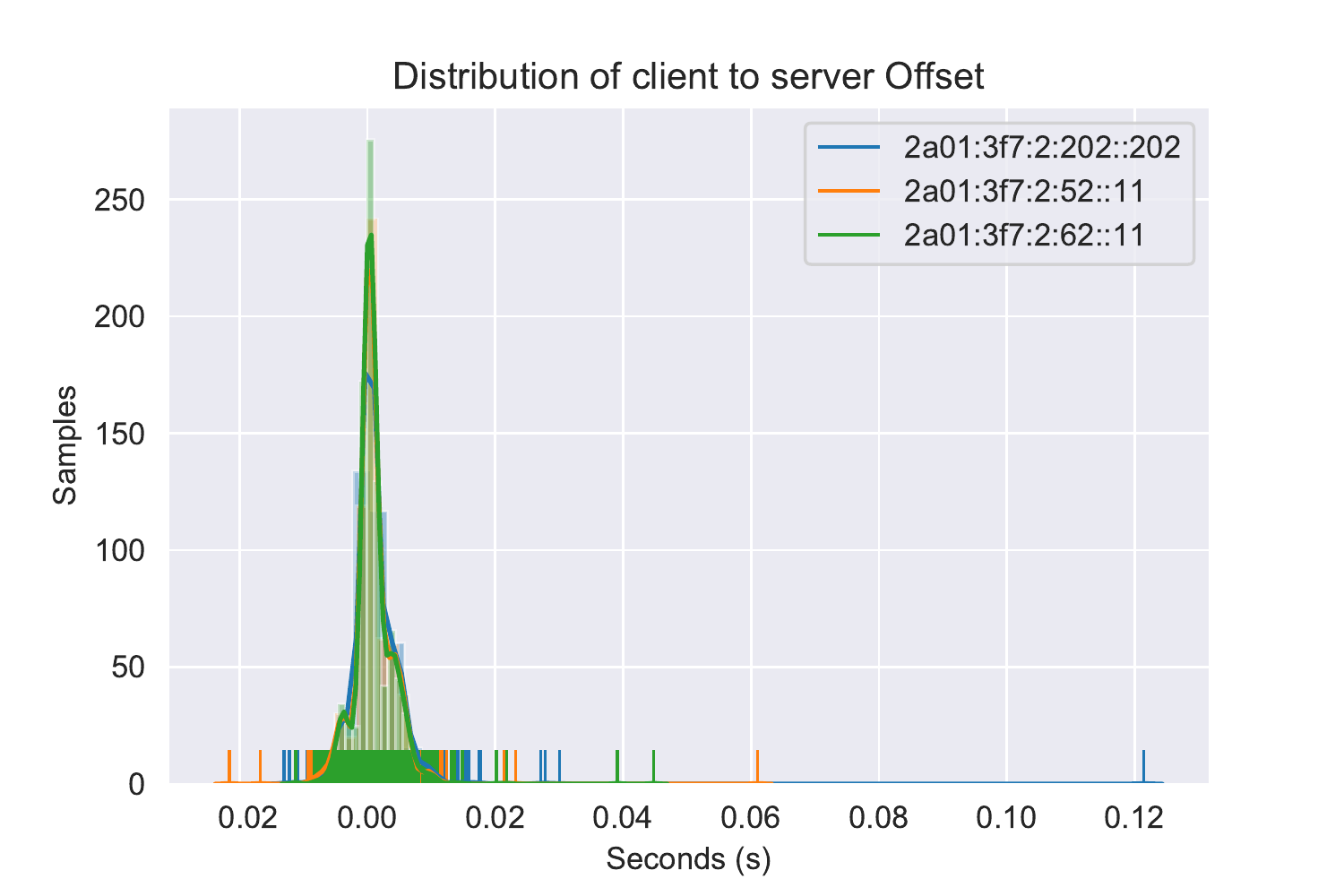}
		\caption{Histogram of NTS Offset distribution for three NTS servers in Sweden. Measurement details are provided in Table \ref{table:ntsperf}}
	\end{subfigure}
	\caption{NTS performance plots}
	\label{figure:ntpplot}
\end{figure*}

\begin{figure*}[!t]
	\centering
	\begin{subfigure}{0.49\textwidth}
		\centering
		\includegraphics[height=2.5in]{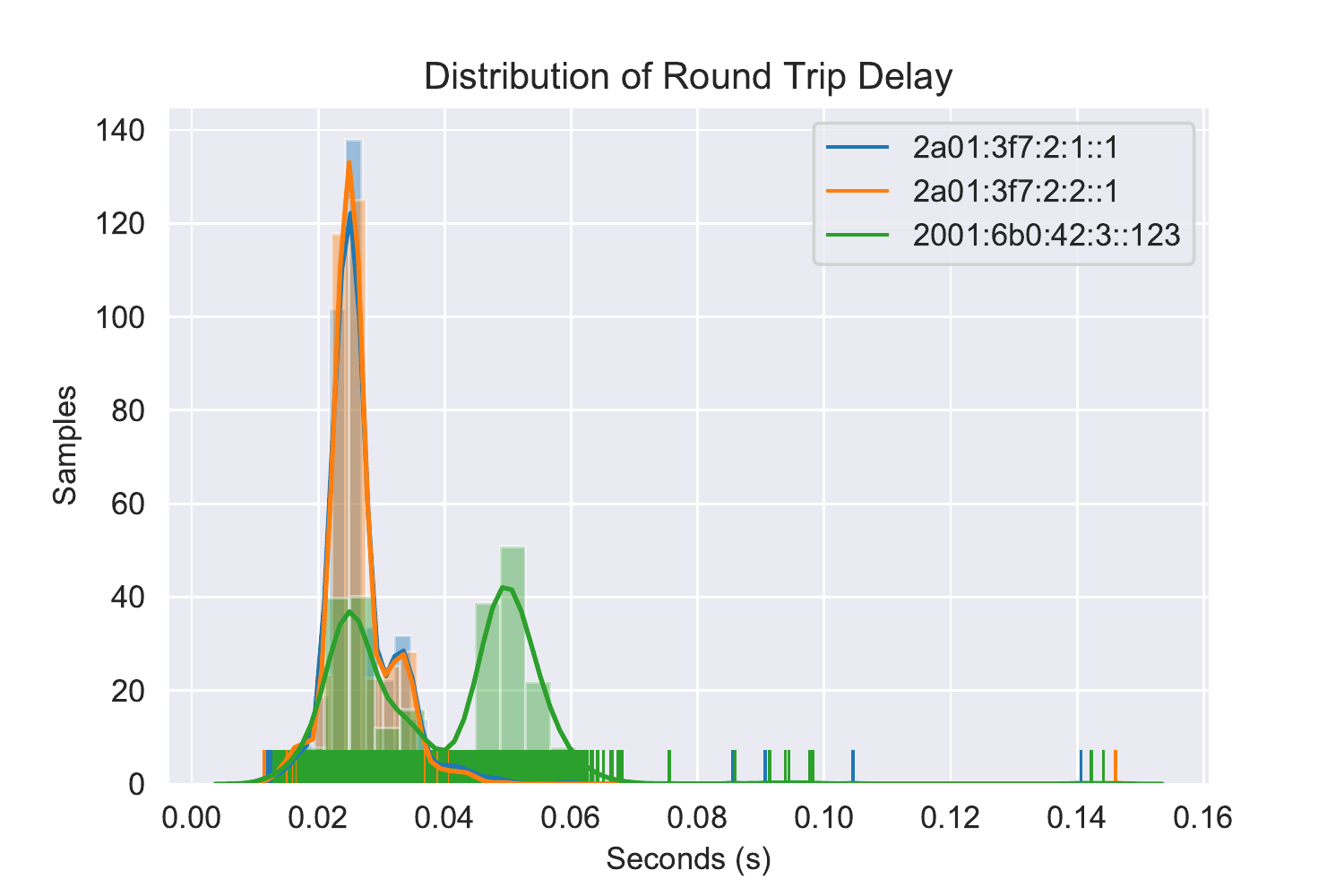}
		\caption{Histogram of NTP Round Trip Delay distribution for three NTP servers in Sweden. Measurement details are provided in Table \ref{table:ntpperf}} 
	\end{subfigure}
	~
	\begin{subfigure}{0.49\textwidth}
		\centering
		\includegraphics[height=2.5in]{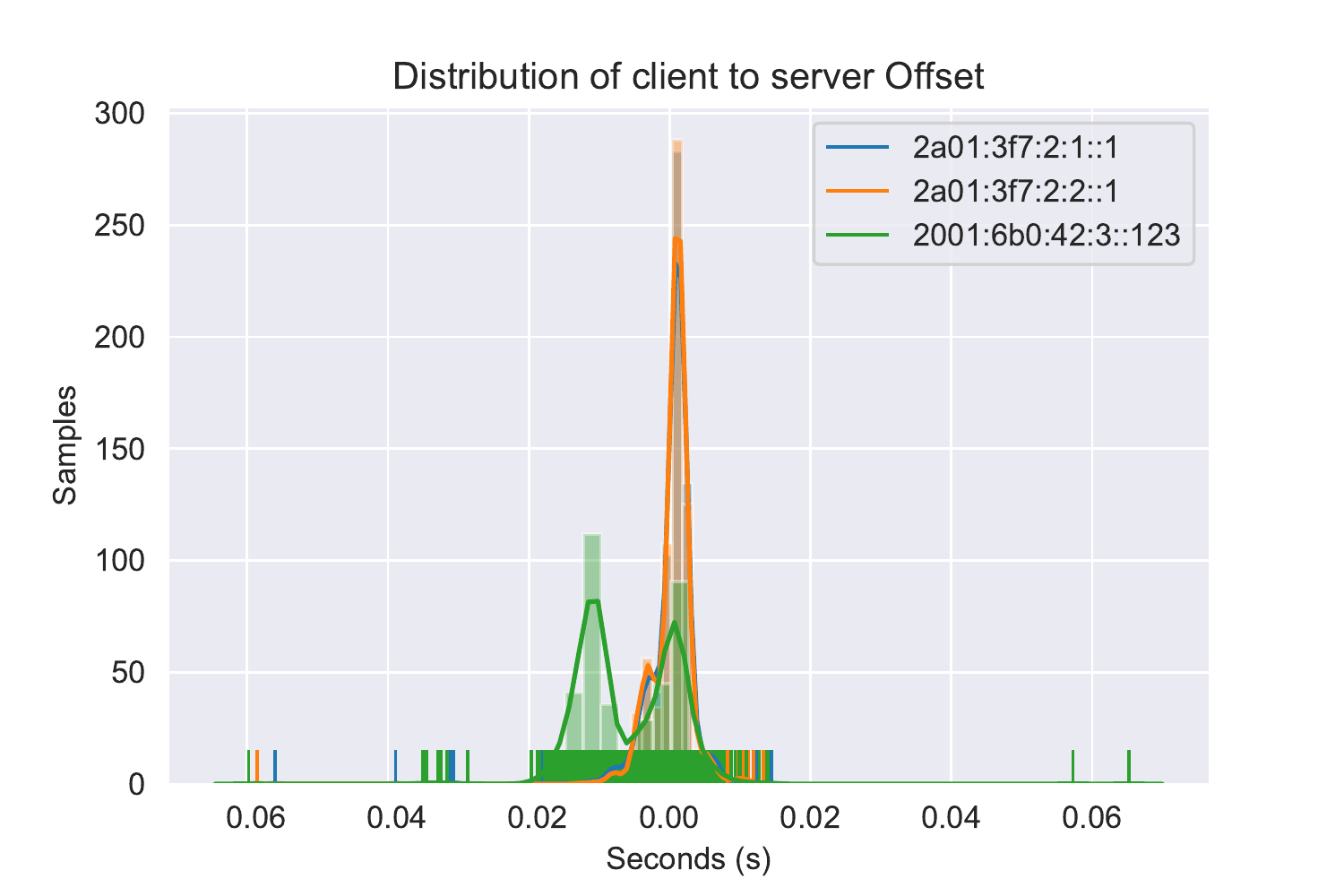}
		\caption{Histogram of NTP Offset distribution for three NTP servers in Sweden. Measurement details are provided in Table \ref{table:ntpperf}}
	\end{subfigure}
	\caption{NTP performance plots}
	\label{figure:ntsplot}
\end{figure*}

The results presented for NTS and NTP pertain to clients (mobile nodes) connected to the Internet for extended periods of time. If only short lived connections are available, the achievable accuracy depends on the local clock stability. Intuitively, when a remote time source is not available, the mobile client can still use its local clock. If a remote time source is available at any time, it can be used to continuously correct the local clock to provide a stable reference. If the remote time source is intermittently available, the local clock will drift depending on its quality.

The security improvements obtained from NTS come at an overhead due to the additional operations required by the NTS protocol. The setup of the cryptographic keys introduces an initial overhead, even if it is a one-time operation. The Message Authentication Codes computation for each NTS transaction introduces limited overhead. According to the analysis in \cite{nts-performance}, based on a previous version of the NTS standard, NTS generates approximately six times more computational load compared to NTP, for the same number of transmitted and received packages. On the other hand, only a fraction of this overhead is caused by the additional authentication (approximately 3\%). In the context of this work, the computational overhead is one of the considered aspects. The accuracy of NTS compared to NTP and the increased level of robustness still make NTS a preferable solution for GNSS spoofing detection. Additionally, the polling cycle can be regulated to limit the overhead caused by the NTS protocol.

Considering the proposed extension for Wi-Fi, the scheduling of the beacons for transmission preserves the time information accuracy. To achieve accuracy within the standard's tolerance, the beacon scheduling system must be implemented at the driver level; while the pre-computation and assembly of the payloads can be performed at the application level.

According to the 802.11 physical layer specification, each beacon is transmitter at a specific Target Beacon Transmission Time (TBTT). If the medium is available at the requested TBTT (in out case 100TU), the scheduled beacon is transmitted. If the medium is not available at the requested TBTT, it will be accessed as per the 802.11 physical layer specification (section 11.1.2.1). In this case the beacon may be delayed because of the Carrier Sense Multiple Access (CSMA) deferral. All the subsequent beacon transmissions will be scheduled at the nominal TBTT. 

The standard (section 11.1.2.4) mandates a minimum accuracy of the Time Synchronization Function (TSF) timer to be 100ppm. This limits the accuracy of the synchronization between STA and AP. Specifically, for out application, it limits the accuracy of the timestamps in the beacons. The synchronization accuracy of $\pm 0.01\%$ over one TBTT of 100TU is within \SI{25}{\micro \second}, accounting for other delays due to the transmission in the physical layer.

The current setup is based on a software simulator able to  provide a realistic beacon distribution, with emission interval (TBTT and TSF synchronization error) standard deviation comparable to an hardware transmitter. This effect is currently simulated by adding a 100ppm error to the beacon timestamps. Nevertheless, the current implementation shows the possibility of achieving authentication with low overhead, given the rate of the beacons. Preliminary measurements show that the results in terms of accuracy (and in terms of effectiveness in detecting the GNSS attacks) will be very close to the ones presented in \cite{kzmsppPLANS2020}, where the implementation had no security features. 

\begin{figure}
	\centering
	\begin{subfigure}[t]{\linewidth}
		\includegraphics[width=\linewidth]{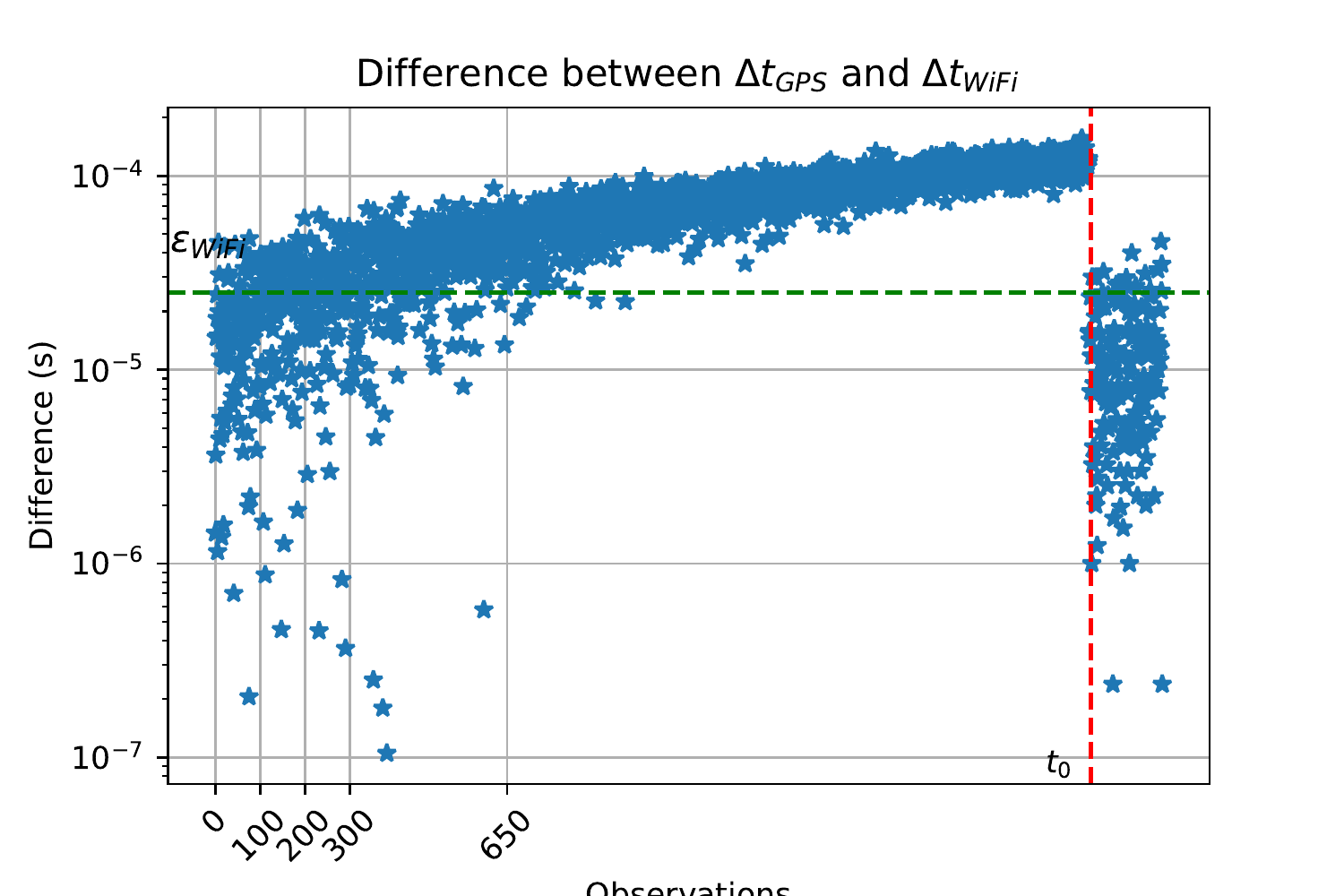}
		\caption{Observation window of \SI{1}{\second}}
	\end{subfigure}
	
	\begin{subfigure}[t]{\linewidth}
		\includegraphics[width=\linewidth]{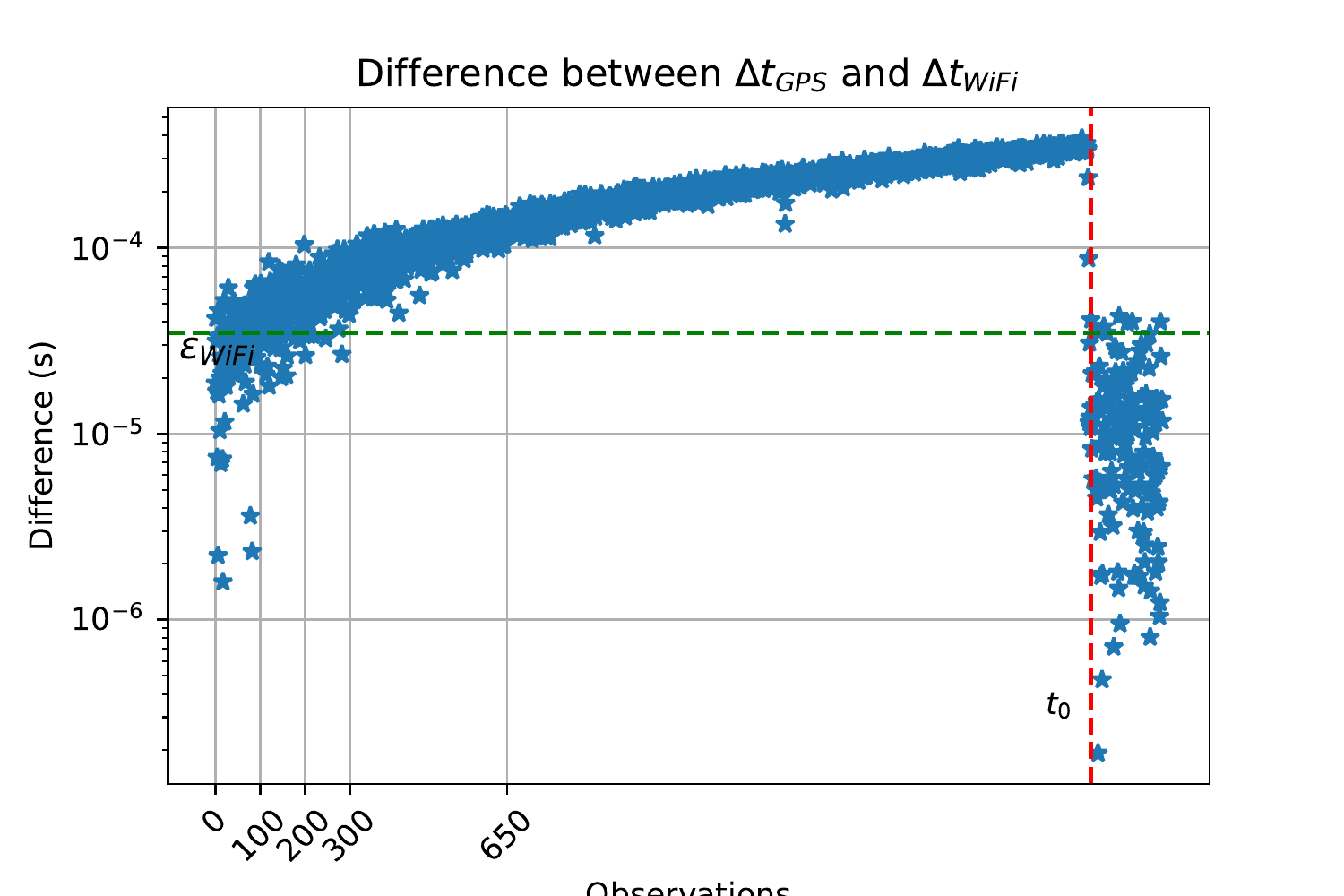}
		\caption{Observation window of \SI{3}{\second}}
	\end{subfigure}
	
	\begin{subfigure}[t]{\linewidth}
		\includegraphics[width=\linewidth]{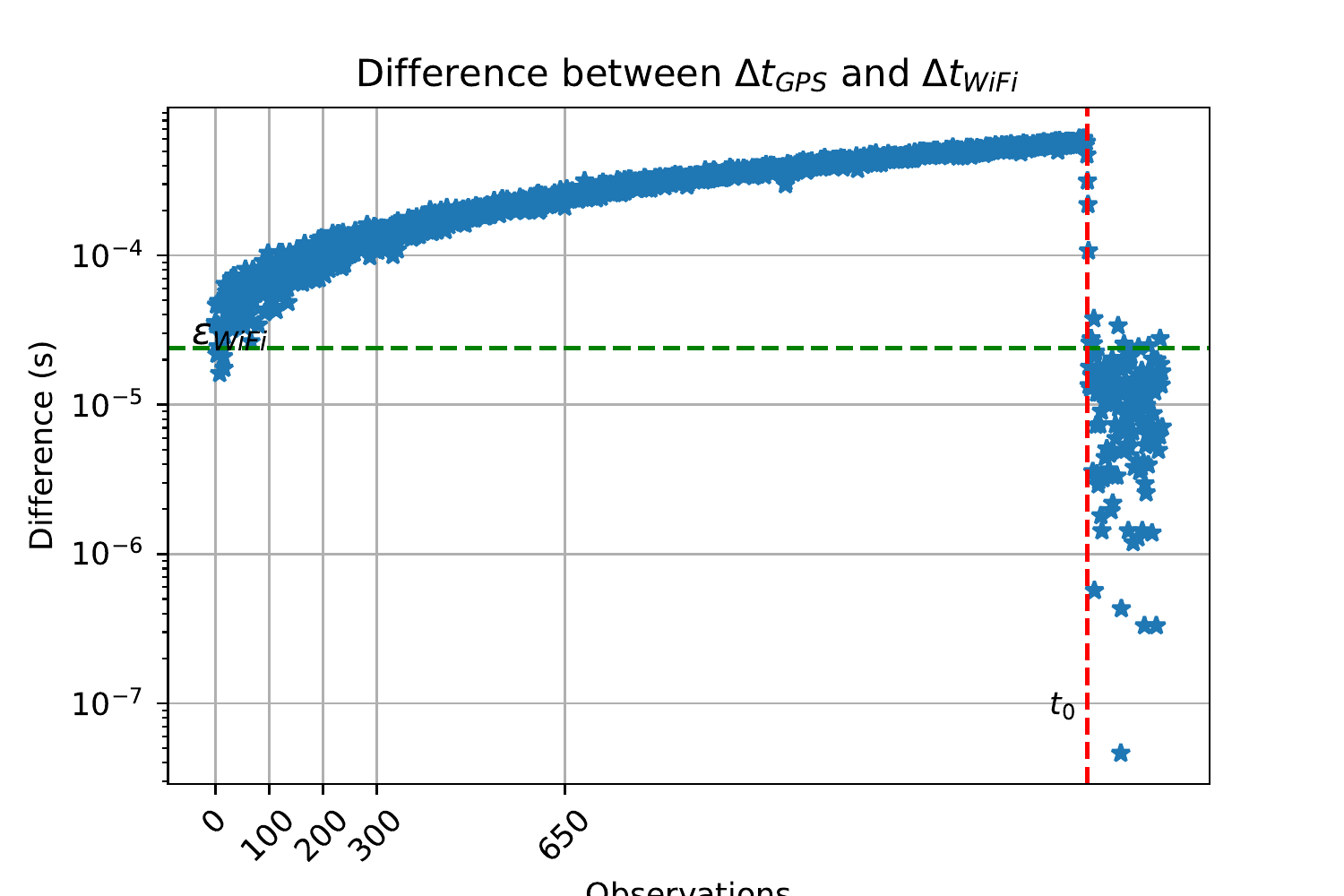}
		\caption{Observation window of \SI{5}{\second}}
	\end{subfigure}
	\caption{GNSS spoofing detection with authenticated beacon timestamps. Results based on a simulated transmitter. The accuracy of the detection system improves with longer observation windows}
	\label{fig:sim-results}
\end{figure}

Similarly to the measurements performed in \cite{kzmsppPLANS2020}, the GNSS time is modified by progressively increasing time offsets until a total bias on the real GNSS time of \SI{120}{\milli \second} is obtained at time $t_0$. After $t_0$, the time bias is kept constant. A detection threshold based on the beacon timestamp accuracy is defined ($\epsilon_{Wi-Fi} = \SI{25,5}{\micro \second}$, obtained from the simulated beacon transmitter), given by the standard deviation of the received beacon timestamps. Different observation windows are identified, corresponding to increasing beacon intervals ($T_{Window} = {1,3,5}\si{\second}$). Figure ~\ref{fig:sim-results} shows the GNSS spoofing detection capability of the present implementation of the setup, with different observation windows. Only beacons that pass the authentication test are used to detect potential attackers. 

We briefly discuss the computational cost of the proposed Wi-Fi augmentation scheme. The cost of an HMAC operation is low in modern platforms, considerably lower than verifying a digital signature. For a standard Intel i7 laptop CPU, there is one order of magnitude difference in the delay between calculating one HMAC and verifying one signature. 
The amortization of the few signature verifications over a large number of beacons reduces the cost of verification, by reducing the number of the most computationally expensive operation. Table \ref{table:beacon-cost} presents the typical delay of each cryptographic operation, when performed on an Intel i7-8750H mobile CPU.

As cost metric, we consider the time spent by the client to authenticate the beacons time information. We define $C_{SIG}$, cost of validating one digital signature, $C_{HMAC}$, cost of calculating one HMAC and $C_{HASH}$, cost of computing one hash element.
If the receiver does not have any prior information on the state of the AP, the cost of the first authentication is calculated as $C^\prime = C_{SIG} + C_{HMAC} + k \cdot C_{HASH}$; plus the cost to validate the AP certificate if needed.
The cost of first verification increases linearly with the distance $k$ of the current hash element element from the most recent digitally signed anchor, but it is bounded by the frequency of the signed anchors. In the worst case scenario, the client is not able to retrieve any digitally signed anchor and has to compute the entirety of the $H_C$. All the subsequent authentications have a cost $C^{\prime \prime} = C_{HMAC} + C_{HASH}$; or multiple $C_{HASH}$ in case of loss of some beacons.
Clients that frequently transition between coverage areas of different APs are subject to a first verification every time they enter a new AP area.

\begin{table}[h!]
\centering
 \begin{tabularx}{\columnwidth}{||X|X|X|X||} 
 \hline
 Operation & Op \# & Avg. Delay & Total\\
 \hline
 Signature\_verify (RSA1024) & 1 & \SI{52}{\micro \second} & \SI{52}{\micro \second} \\
 \hline
 HMAC (SHA256) &  & \SI{7.67}{\micro \second} & \SI{7.67}{\micro \second} \\
 \hline
 HASH (SHA256) & $k$ & \SI{1.64}{\micro \second} & $k \cdot$\SI{1.64}{\micro \second} \\
 \hline
 \end{tabularx}
\caption{Cost of cryptographic operations for authenticating Wi-Fi beacons at the receiver. Computations are performed on an Intel i7 cpu.}
\label{table:beacon-cost}
\end{table}

For example, consider an urban scenario, with the client moving at an average speed of \SI{15}{\kilo\meter\per\hour}, and a trajectory of \SI{100}{\meter} within the coverage area of an AP. The total transit time for the AP coverage is 24.03\si{\second} and during this period the total number of beacons a client can receive by one AP is $\approx$ 234\si{\bcn}. 
For intermediate signed anchor periods longer than the travel time of the client within the AP coverage area (e.g., the signed anchor is updated every \SI{60}{\second}, corresponding roughly to \SI{600}{\bcn}), the client needs to verify on average only one anchor for every new AP, with a worst case of $k=599$. In this case, the time spent for authenticating the beacons is $\approx \SI{3,21}{\milli \second}$, corresponding to $0.013\%$ of the transit time.

For APs with hash chains valid for long periods of time (i.e., one hash chain valid for \SI{24}{\hour}), the usage of intermediate anchors effectively reduce the cost of joining the coverage area of an AP. Without intermediate anchors, a client joining the coverage area close to the hash chain expiration time would be penalized. The distribution of the intermediate anchors can be performed directly by the AP, at regular intervals, or through the Internet. The former solution requires larger beacon payloads, while the latter supposes (at least intermittent) network connectivity.

\section{Conclusions}
\label{section:conclusion}
Alternative sources of time are beneficial in detecting time-manipulating GNSS spoofers. Opportunistic time providers, Wi-Fi beacons, and dedicated timing infrastructures, NTP, provide largely available, precise sources of time information. While proving effective in detecting misbehaving GNSS receivers, i.e., when the client is able to obtain enough information, unauthenticated time information can be relied upon.

Different options to secure augmentation time information are investigated, notably Network Time Security (NTS) and modified Wi-Fi beacons to support authentication. This scheme imposes limited overhead, does not disrupt the normal operation of the Wi-Fi APs and can be deployed readily.

Furthermore, considering NTP and NTS, there is no significant performance loss due to the authentication of the NTP frames, making NTS a valid solution to provide GNSS-based receivers secure Internet time.

\section{Acknowledgment}
This work supported by the Swedish Foundation for Strategic Research (SSF) SURPRISE project and the KAW Academy Fellowship Trustworthy IoT project. In addition, we would like to thank Dr. Hongyu Jin for the constructive feedback and discussion during the development of this work.

\bibliographystyle{IEEEtran}
\bibliography{ref}

\end{document}